\documentclass[a4paper,fleqn]{cas-dc}

\usepackage[numbers,sort&compress]{natbib}

\usepackage{caption}
\usepackage[none]{hyphenat}
\usepackage{wasysym} %for chechek
\usepackage{multirow,hhline}

\usepackage{dsfont} %for equation form like map model
\usepackage{array} %for ticker horizental line in table
\newcolumntype{?}{!{\vrule width 1pt}} %for ticker vertical line in table
\usepackage{bm} %to write bold font in math mode
\usepackage{makecell} %to draw thick line in tables
\usepackage{endnotes}
\let\footnote=\endnote
\definecolor{myblue}{RGB}{216, 253, 255}
\definecolor{mybrown}{rgb}{0.5, 0.0, 0.0}
\definecolor{blueTitle}{rgb}{0.0, 0.22, 0.66}
\usepackage{adjustbox} %for table resize
\usepackage[linesnumbered,lined,ruled]{algorithm2e}
\let\oldnl\nl% Store \nl in \oldnl
\newcommand{\nonl}{\renewcommand{\nl}{\let\nl\oldnl}}% Remove line number for one line
\usepackage{pifont} %to add sepcial charcter for symbol of itemize
\begin{document}
\let\WriteBookmarks\relax
\def\floatpagepagefraction{1}
\def\textpagefraction{.001}
\shorttitle{A Survey on Reproducible Domain-Specific Knowledge Graphs}
\shortauthors{Babalou et al.}

\title [mode = title]{Reproducible Domain-Specific Knowledge Graphs in the Life Sciences: a Systematic Literature Review}
%\tnotemark[1,2]

%\tnotetext[1]{This document is the results of the research
%   project funded by the National Science Foundation.}
%
%\tnotetext[2]{The second title footnote which is a longer text matter
%   to fill through the whole text width and overflow into
%   another line in the footnotes area of the first page.}

\author[1,2]{Samira Babalou}[bioid=1,
                        orcid=0000-0002-4203-1329]
\cormark[1]
%\fnmark[1]
\ead{samira.babalou@uni-jena.de}
%\ead[url]{www.cvr.cc, cvr@sayahna.org}

\credit{Conceptualization of this study, existing Knowledge Graphs analysis, Original draft preparation}

\address[1]{Heinz Nixdorf Chair for Distributed Information Systems,
	Institute for Computer Science, Friedrich Schiller University Jena, Germany}
\address[2]{German Center for Integrative Biodiversity Research (iDiv), Halle-Jena-Leipzig, Germany}

\author[1,3]{Sheeba Samuel}[bioid=1,
orcid=0000-0002-7981-8504]
%\cormark[1]
%\fnmark[1]
\ead{sheeba.samuel@uni-jena.de}
%\ead[url]{www.cvr.cc, cvr@sayahna.org}

\credit{Conceptualization of this study, existing Knowledge Graphs analysis, Original draft preparation}

%\address[1]{Heinz-Nixdorf Chair for Distributed Information Systems, Institute for Computer Science, Friedrich Schiller University Jena, Germany}
%Leutragraben 1, Jentower,  Room 17N05,  07743 Jena, Germany}
%\address[2]{Michael Stifel Center Jena, Jena, Germany}

\author[1,2,3]{Birgitta K\"{o}nig-Ries}[bioid=1,
orcid=0000-0002-2382-9722]
\ead{birgitta.koenig-ries@uni-jena.de}
%\ead[URL]{www.sayahna.org}

\credit{Supervision, Validation, review \& editing}

\address[3]{Michael Stifel Center Jena}

\cortext[cor1]{Corresponding author; Address: Leutragraben 1, Jentower,  Room 17N05,  07743 Jena, Germany}
%\cortext[cor2]{Principal corresponding author}
%\fntext[fn1]{This is the first author footnote. but is common to third
%  author as well.}
%\fntext[fn2]{Another author footnote, this is a very long footnote and
%  it should be a really long footnote. But this footnote is not yet
%  sufficiently long enough to make two lines of footnote text.}

%\nonumnote{This note has no numbers. In this work we demonstrate $a_b$
%  the formation Y\_1 of a new type of polariton on the interface
%  between a cuprous oxide slab and a polystyrene micro-sphere placed
%  on the slab.
%  }

\begin{sloppypar}
\begin{abstract}
Knowledge graphs (KGs) are widely used for representing and organizing structured knowledge in diverse domains.
However, the creation and upkeep of KGs pose substantial challenges. 
Developing a KG demands extensive expertise in data modeling, ontology design, and data curation. 
Furthermore, KGs are dynamic, requiring continuous updates and quality control to ensure accuracy and relevance. 
These intricacies contribute to the considerable effort required for their development and maintenance.
One critical dimension of KGs that warrants attention is reproducibility. 
The ability to replicate and validate KGs is fundamental for ensuring the trustworthiness and sustainability of the knowledge they represent. 
Reproducible KGs not only support open science by allowing others to build upon existing knowledge but also enhance transparency and reliability in disseminating information.
Despite the growing number of domain-specific KGs, a comprehensive analysis concerning their reproducibility has been lacking.
This paper addresses this gap by offering a general overview of domain-specific KGs and comparing them based on various reproducibility criteria.
Our study over 19 different domains shows only eight out of 250 domain-specific KGs (3.2\%) provide publicly available source code.
Among these, only one system could successfully pass our reproducibility assessment (14.3\%).
These findings highlight the challenges and gaps in achieving reproducibility across domain-specific KGs.
Our finding that only 0.4\% of published domain-specific KGs are reproducible shows a clear need for further research and a shift in cultural practices.
\end{abstract}

%\begin{graphicalabstract}
%\includegraphics{figs/grabs.pdf}
%\end{graphicalabstract}

%\begin{highlights}
%\item Merging multiple ontologies in the same time with a partitioning-based method
%\item Partitioning a set of source ontologies to a group of blocks based on their structure
%\item Comparing the performance of the n-ary merge strategy versus the binary method
%\item An online ontology merging tool which scale to many sources
%\end{highlights}


\maketitle

%Main content

\section{Introduction}
At their core, Knowledge Graphs (KGs) %are widely regarded as one of the most promising ways to manage and link information in the age of Big Data.
are structured information about a particular domain in the form of entities and relations. They are used in different applications such as recommendation systems~\cite{wu2020event}, health misinformation detection~\cite{cui2020deterrent}, or disease characteristics identification~\cite{zhu2020integrative}. 
While different definitions of KGs exist, we use the definition provided by Hogan et al.~\cite{hogan2021KG}. According to their definition, KGs are graph of data intended to accumulate and convey knowledge of the real world, whose nodes represent entities of interest and whose edges represent potentially different relations between these entities.

%While the definition of KGs can be ambiguous \cite{hogan2021KG, nickel2015review}, we use the definition provided by Nickel et al.~\cite{nickel2015review}. According to their definition, KGs are graph-structured knowledge bases that store factual information through relationships between entities. These graphs consist of nodes representing concepts and real-world entities, and edges denoting connections between entities. Both nodes and edges carry unique identifiers. \sheeba{I added the hogan2021KG paper and would use the KG definition from this paper because it is introduced by the semantic web community}

%A Knowledge Graph (KG) is a graph of data intended to accumulate and convey knowledge of the real world, whose nodes represent entities of interest and whose edges represent potentially different relations between these entities~\cite{hogan2021KG}. 

Different definitions of the term reproducibility exist~\cite{taylor1994,result2017,plesser2018reproducibility,national2019reproducibility,acm2020artifact,samuel2021understanding}.
According to~\cite{taylor1994,result2017,samuel2021understanding}, reproducibility is the capability of getting the same (or close-by) results whenever an experiment is carried out by an independent experimenter using different conditions of measurement which include the method, location, or time of measurement.
Reproducibility is defined as obtaining consistent computational results using the same input data, steps, methods, code, and analysis conditions, according to \cite{national2019reproducibility,acm2020artifact}.
The importance of achieving reproducibility is underlined by the many studies and surveys that have been done to check the reproducibility of published results in different fields \cite{ioannidis2009repeatability,begley2013,baker20161500,raff2019}.
More significant insights were brought into the reproducibility crisis by the survey conducted by Nature in 2016 \cite{baker20161500}.
The difficulty in reproducing published results can also be seen in computational science \cite{raff2019,pimental2019,samuel2023computational}.
These works indicate the continued existence of a problem in reproducing published results in different disciplines.

%TODO: For the following parargraph, Birgitta asked to re-write that like: "Unclear, why this is important. I would strucutre this paragraph as follows. KGs are useful (eg. greater alingment.... ). KGs should be repdorucible, because :a that lowers effort and neeed expertise, and b increases trusts.... In other fields, where reproducibility is also desirable studies show, that it is ofen lackingW- With this paper, we aim to investaigate whether thiis is also an issue for KGs.
KGs can facilitate a greater alignment between data and expertise in every domain, making data more accessible and usable. Although KGs are useful in various domains, their broad uptake is still hindered by the substantial effort and the high semantic web expertise needed to create them.
%TODO: here we should add more why reproducbility is importatn
While reproducibility should be a standard practice in scientific endeavors, most existing KGs do not offer the ability to recreate or reproduce them.
%TODO; for the last sentence, Birgitta commented that: "This is already the outcome of the paper." but I do not know whether to re-write or move it somewhere else?!
Despite increased awareness of the problem and the rising availability of data and code used in publications, reproducing published results remains challenging. This holds true for KGs as well. However, a reproducible KG can foster trust in the information provided and support open data and open science practices.
%This is an important step towards strengthening the establishment of open science practices in different domains.
The significance of ensuring reproducibility in knowledge graph generation has been highlighted in~\cite{van2021leveraging}.

%these paper might be interesting~\cite{zhang2019vision,yu2020domain,qi2020scratch,fan2017dkgbuilder}

%TODO: should we have a separate section for the literature review (survey on the existing KG-Survey) or this paragraph is enough?
With the popularity of KGs, nowadays, many KGs are generated for different domains and in various applications. Accordingly, numerous researchers surveyed the existing KGs exploring multiple aspects such as embeddings (cf.~\cite{gesese2019survey,lu2020utilizing,wang2021survey,dai2020survey}), refinements~\cite{paulheim2017knowledge}, applications~\cite{zou2020survey}, architectures~\cite {zhao2018architecture}, privacy-preservation~\cite {chen2020survey}, completion (cf.~\cite {nguyen2017survey,arora2020survey}, question answering~\cite{yani2021challenges}, among others.

To the best of our knowledge, the only research on surveying domain-specific KGs was introduced by Abu-Salih in~\cite{abu2021domain}, which differs from our study as we specifically focus on the reproducibility aspects of KGs.
In this paper, we take the first step towards analyzing the existing KGs with respect to their reproducibility. We first provide an overview of the existing domain-specific KGs and compare them based on general criteria, including the respective domain, resource type, and construction method. This comparative analysis gives readers more insights into the existing domain-specific KGs. We then investigate the extent to which the KGs are reproducible using a defined set of criteria that reflect the reproducibility aspect. In this paper, we attempt to reproduce knowledge graphs using the same data and methods provided by the original authors in an experimental setup closely resembling theirs.

Although the main focus of this study is the reproducibility of existing domain-specific KGs, it is worth noting that the aspects of findability, accessibility, and interoperability, as emphasized by the FAIR principles~\cite{wilkinson2016fair}, constitute an interesting research direction. However, analyzing these aspects is beyond the scope of the current study and could be a potential avenue for future research.

The remainder of this paper is structured as follows: Section~\ref{sec:surveyMethodology} shows the survey methodology. Section~\ref{sec:existingKGs} presents the existing domain-specific Knowledge Graphs and the criteria for their reproducibility, followed by the discussion in Section~\ref{sec:discussion}. The conclusion and future works are presented in Section~\ref{sec:conclusion}.

%\section{Preliminaries} \label{sec:preliminaries}
%TODO: do we need this section?

\section{Survey Methodology} \label{sec:surveyMethodology}
We first searched for the keyword ``domain knowledge graph'' in the Google Scholar search engine\footnote{\url{https://scholar.google.de/} accessed on 17.01.2022}. %\sheeba{Was it on 2022 or 2021?, I guess 2021}. 
We limited our search to papers published until the end of 2021.
At the time of querying (Jan 01, 2022), this search resulted in 713 papers. We looked at their domain names (e.g., biodiversity, geoscience, biomedical, etc.) and then extended our search for those specific domain names that appear on the first result, e.g., for ``biodiversity knowledge graph'', ``biomedical knowledge graph'', and so on.
To ensure the exclusion of duplicate entries for the "domain knowledge graph" that may have appeared in multiple categories, we removed such duplicates. As a result, we identified a collection of 603 unique papers focused on the "domain knowledge graph."
Overall, our research encompassed a total of 1759 papers across 19 distinct domains.
Note that we excluded the paper by Kim~\cite{kim2021knowledge} from our analysis as we were unable to access and ascertain whether it pertained to KG creation, despite attempts to contact the author. %It comes in the medical group

We have selected a subset of the papers listed in the search results by considering these criteria: (i) we chose articles written in English only, (ii) we selected papers that focused on the creation or construction of knowledge graphs (KGs). Papers that primarily addressed the usage or other aspects of KGs were excluded. Moreover, the search results from Google Scholar displayed papers where the keywords appeared in the title, introduction, or state-of-the-art sections. Some papers do not focus on the topic of our keywords. However, some papers only briefly mentioned the keywords in the state of the art, indicating that they did not primarily focus on generating KGs. Therefore, we disregarded such papers.
The selection process was carried out manually, thoroughly examining each paper to determine its relevance to KG construction. As a result, out of the initial 1759 papers listed in Google Scholar, we identified 250 papers that met our selection criteria.

From this subset, we further narrowed down our selection to papers that provided open-source code. We checked all 250 papers manually by looking at the paper content, whether they have a link to the GitHub repository or any web pages where their code is published. We also checked the data availability statement section in papers, if available.
%we searched for GITHUB, AVAILABLE, WWW, HTTP, LINK, OPEN SOURCE terms
Surprisingly, we only found eight papers out of 250 with open-source code.

We use a script to download the articles to ensure the reproducibility of our experimental results. The script, the original search results obtained from Google Scholar, and our analysis of the results (whether each paper is selected or not, and whether they are open-source or not) are published in our repository\footnote{\url{https://github.com/fusion-jena/iKNOW/tree/main/Reproducibility-Survey}}.

Table~\ref{table:keywordSearch} presents a summary of our keyword search results, indicating the number of published papers found on each respective topic as retrieved from Google Scholar. The third column shows the number of papers on Google Scholar for each keyword. The fourth column specifies the count of selected papers relevant to Knowledge Graph (KG) construction, while the final column denotes the number of papers accompanied by open-source code. The last row of this table shows the total number of papers for each category.

\begin{table}[bt!]
	\caption{Keyword search on the Google Scholar. |Papers| denotes the total number of papers retrieved for a given keyword; |Selected| shows the number of selected papers related to building KGs; |Open-source code| shows the number of papers that provides open-source code.}
	\centering
	\resizebox{8.5cm}{!} {
	\begin{tabular}{ |c|c|c|c|c| }
		\hline
		\rowcolor{gray!15} & \textbf{Keyword} & & & \textbf{|Open-} \\
		\rowcolor{gray!15}\multirow{-2}{*}{\textbf{no.}} & \textbf{search} & \multirow{-2}{*}{\textbf{|Papers|}} & \multirow{-2}{*}{\textbf{|Selected|}}  & \textbf{source|}\\  		[0.5ex]	\Xhline{3\arrayrulewidth}
		1 & ``Domain knowledge graph'' & 602 & 88 & 2 \\ \hline
		\rowcolor{gray!15}2 & ``Agriculture knowledge graph'' & 16 & 5 & 0 \\ \hline
		3 & ``Biodiversity knowledge graph'' & 87 & 5 & 1\\ \hline
		\rowcolor{gray!15}4 & ``Biomedical knowledge graph'' & 214 & 12 & 2\\ \hline
		5 & ``Cultural knowledge graph'' & 17 & 6 & 0\\ \hline
		\rowcolor{gray!15}6 & ``E-commerce knowledge graph'' & 16 & 7 & 0 \\ \hline
		7 & ``Education knowledge graph'' & 32 & 14& 0\\ \hline
		\rowcolor{gray!15}8 & ``Financial knowledge graph'' & 64 & 3 & 0\\ \hline
		9 & ``Geographic knowledge graph'' & 117 & 20 & 1 \\ \hline
		\rowcolor{gray!15}10 & ``Geoscience knowledge graph'' & 9 & 4 & 1 \\ \hline
		11 & ``Healthcare knowledge graph'' & 45 & 5 & 0 \\ \hline
		\rowcolor{gray!15}12 & ``Industrial knowledge graph'' & 37 & 8 & 0 \\ \hline
		13 & ``Medical knowledge graph'' & 291 & 38 & 0 \\ \hline
		\rowcolor{gray!15}14 & ``Military knowledge graph'' & 26 & 8 & 0\\ \hline
		15 & ``Movie knowledge graph'' & 48 & 6 & 0 \\ \hline
		\rowcolor{gray!15}16 & ``Political knowledge graph'' & 6 & 0 & 0 \\ \hline
		17 & ``Robotic knowledge graph'' & 4 & 1 & 0\\ \hline
		\rowcolor{gray!15}18 & ``Security knowledge graph'' & 80 & 10 & 0\\ \hline
		19 & ``Tourism knowledge graph'' & 42 & 9 & 1 \\ \hline
		\rowcolor{gray!15}20 & ``Water knowledge graph'' & 3 & 1 &0\\ \hline \hline
		\multicolumn{2}{|c|}{Total} & 1756 & 250 & 8\\ \hline
	\end{tabular}
}
\label{table:keywordSearch}
\end{table}

\section{Reproducibility of domain-specific Knowledge Graphs} \label{sec:existingKGs}
This paper centers its focus on the aspect of reproducibility. Consequently, as an initial step, we scrutinized all the selected papers to determine the availability of publicly accessible code for the Knowledge Graphs (KGs) they developed. It emerged that only eight papers out of the total 250 (3.2\%) met this criterion.
Note that AliCG (Alibaba Conceptual Graph)~\cite{Alicg}\footnote{\url{https://github.com/alibaba-research/ConceptGraph}} and the KG proposed by Hoa et al.,~\cite{hao2021construction} (for surveying and remote-sensing applications)\footnote{\url{https://github.com/hao1661282457/Knowledge-graphs}}, published only the raw data and not the code. So, these papers were not considered within the category of open-source code.
Moreover, in the biomedical domain, we found two different publications~\cite{dougan2020,dougan2021} related to CROssBAR-KG. We consider them as one unique KG for our further analysis.
%TODO:Survey~ \cite{buchgeher2020knowledge}

In this section, we first summarize the domain-specific KGs that provide open-source code. We then provide a general overview of them in subsection~\ref{subsection-comp-KG} and discuss their reproducibility aspect in Subsection~\ref{subsection-comp-reproduc}.
Existing KGs with open-source code:

\begin{itemize}
	%\item \textbf{AliCG}~\cite{Alicg} (Alibaba Conceptual Graph) is a

	\item \textbf{CKGG}~\cite{CKGG} (Chinese Knowledge Graph for Geography) is a KG covering the core geographical knowledge at the high-school level, containing 1.5 billion triples.
	The authors used a variety of NLP tools to integrate various kinds of geographical data in different formats from diverse sources (such as GeoNames\footnote{\url{https://www.geonames.org/.}}, Wikipedia).
	They conducted a preliminary evaluation of CKGG and showed a prototype educational information system based on CKGG.

	\item \textbf{CROssBAR-KG}~\cite{dougan2020,dougan2021} Knowledge graph presents biological terms as nodes and their known or predicted pairwise relationships as edges. They are directly obtained from their integrated large-scale database, built upon a set of biomedical data resources. The data is enriched with a deep-learning-based prediction of relations between numerous biomedical entities.
	At first, the data is stored in a non-relational database. Then, biologically relevant small-scale knowledge graphs are constructed on the fly, triggered by users' queries with a single or multiple term(s). The system is tested by a use-case study of the COVID-19 dataset.
	%reasoning?: additional relation types are incorporated to the graph as edges between the existing nodes to further enrich the provided information....inferring the missing relations between existing data points
	% aim: aid biomedical research, especially to infer mechanisms of diseases in relation to biomolecules, systems, and candidate drugs.

	%\item \textbf{Dong et al.,}~\cite{dong}

	\item \textbf{ETKG} (Event-centric Tourism Knowledge Graph)~\cite{wu2020event} is a KG to model the temporal and spatial dynamics of tourist trips. %or- which integrates tourism events and their temporal relations to represent tourists' activities %to interconnect events using temporal relations
	The authors extracted information from over 18000 travel notes (structured and unstructured information) crawled from the Internet, and defined an ETKG schema to model tourism-related events and their key properties. The schema of ETKG is built upon the Simple Event Model~\cite{van2011design} with augmented properties and classes. The authors constructed an ETKG of Hainan and realized an application of POI recommendation based on it. %events= tourism activities

	\item \textbf{FarsBase}~\cite{farsbase} is a cross-domain knowledge graph in the Farsi language, consisting of more than 500K entities and 7 million relations. Its data is extracted from the Farsi edition of Wikipedia in addition to its structured data, such as infoboxes and tables.
	To build Farsi Knowledge Graph (FKG), the authors first developed an ontology retrieved from DBpedia ontology, based on resources from Farsi Wikipedia.
	Then, they mapped Wikipedia templates to their built ontology.
	They consider Wikipedia as input of the FKG system. To enhance the performance and flexibility of the knowledge base, they stored data in two-level architecture: a NoSQL database for storing data and metadata, and a triplestore for storing the final data.
	Most entities in the FKG have been linked to DBpedia\footnote{\url{https://www.dbpedia.org/}} and Wikidata\footnote{\url{https://www.wikidata.org}} resources by owl:sameAs property.
	A SPARQL endpoint provides access to the knowledge graph.

	\item \textbf{GAKG} (GeoScience Academic Knowledge Graph)~\cite{GAKG} is a large-scale multimodal academic KG, consisting of more than 68 million triples based on 1.12 million papers published in various geoscience-related journals.
	The entities of GAKG have been extracted under a Human-In-the-Loop framework, using machine reading and information retrieval techniques with manual annotation of geoscientists in the loop. The schema of GAKG consists of 11 concepts connected by 19 relations.
	GAKG is updated regularly and can be queried at the SPARQL query Endpoint.
	It is evaluated using two benchmarks.

	\item \textbf{MDKG}~\cite{fu2020integrated} stands for Microbe-Disease Knowledge Graph and is built by integrating multi-source heterogeneous data from Wikipedia text and other related databases. Through a series of natural language processing, they split the text of Wikipedia pages into sentences. Then, using an existing tool, they perform named entity recognition and relationship extraction on the sentences and obtain the interaction triplets. Afterward, other databases are integrated into their KG. Moreover, they used the representation learning method for knowledge inference and link prediction.

	\item \textbf{Ozymandias}~\cite{ozymandias2019}, a biodiversity knowledge graph, combines scholarly data about the Australian fauna from different sources, including the Atlas of Living Australia\footnote{\url{https://www.ala.org.au}}, the Biodiversity Heritage Library, ORCID\footnote{\url{https://orcid.org}}, and links to external KGs like Wikidata and GBIF\footnote{\url{https://www.gbif.org/what-is-gbif}}.

	\item \textbf{RTX-KG2}~\cite{RTX-KG2} is an open-source software system for building and hosting a web API for querying a biomedical knowledge graph.
	The data from 70 core biomedical knowledge-bases are extracted via a set of Extract-Transform-Load (ETL) modules.
	Its schema is built based on an existing metamodel in the biological domain. %the standard Biolink metamodel: Richard Bruskiewich, Deepak Unni, Chris Mungall, et al. biolink/biolink-model: 2.0.0,	2021. doi:10.5281/ZENODO.4895425.
	%https://biolink.github.io/biolink-model/
	RTX-KG version 2.7.3 contains 10.2 million nodes and 54.0 million edges.

\end{itemize}

\subsection{Comparison of KGs} \label{subsection-comp-KG}
%or General overview of KGs
%In the last section, we briefly explained each KG.
In this section, we summarize the key features of each KG mentioned in Section~\ref{sec:existingKGs}. Table~\ref{table:comparison} shows the comparison of domain-specific KGs with respect to their domain, resource type, construction method, reasoning, cross-linking, evaluation, and year.  The cross-linking aspect indicates whether the elements of the KG are connected to external resources or other KGs such as Wikidata or DBpedia. Note that if the KG is built based on some resources, i.e., the elements of KG are mapped to other data resources, we do not consider them as cross-linking.

\begin{table*}[ht]
	\centering
	\caption{General overview of domain-specific KGs.}
	\begin{adjustbox}{width=1\textwidth}
		\small
		\begin{tabular}{|l|l|l|l|l|l|l|l|}
			\hline
			\rowcolor{gray!15}\textbf{KG}       & \textbf{Domain} & \textbf{Resource Type}            & \textbf{Construction Method} & \textbf{Reasoning} & \textbf{Cross Linking}                 & \textbf{Evaluation} & \textbf{Year}   \\  [0.5ex]	\Xhline{3\arrayrulewidth}

			CKGG & Geography & Data resources & Machine Learning& Not declared& Wikipedia& Yes& 2021 \\ \hline

			\rowcolor{gray!15}CROssBAR-KG      & Biomedical      & Data resources                    & Machine Learning             & Yes                & Not declared                                     & Yes                 & 2020           \\ \hline

			ETKG              & Tourism         & Web pages                         & Machine Learning             & Not declared       & Not declared                                     & Yes                 & 2020            \\ \hline

			\rowcolor{gray!15}FarsBase & Cross-domain & Wikipedia & Heuristic& Not declared& DBpedia, Wikidata & Not provided& 2021 \\ \hline

			GAKG & Geoscience & Publication & Machine Learning& Not declared& Wikidata& Yes & 2021 \\ \hline

			\rowcolor{gray!15}MDKG              & Biomedical      & Wikipedia text                    & Machine Learning              & Yes                & Not declared                                     & Not provided        & 2020           \\ \hline

			Ozymandias         & Biodiversity    & Publication                       & Heuristic                    & Not declared       & Wikidata, GBIF                         & Not provided        & 2019         \\ \hline

			\rowcolor{gray!15}RTX-KG2 & Biomedical & Data resources & Heuristic& Yes & Not declared& Not provided& 2021 \\ \hline

		\end{tabular}
	\end{adjustbox}
	\label{table:comparison}
\end{table*}

\subsection{Criteria for reproducibility of KGs} \label{subsection-comp-reproduc}
Reproducibility is one of the important principles of scientific progress.
It emphasizes that a result obtained by an experiment or observational study should be consistently obtained with a high degree of agreement when different researchers replicate the study with the same methodology. %Only after one or several such successful replications should a result be recognized as scientific knowledge.
%it was copied from wikipedia, we should re-write the sentences above.
Indeed, reproducing an experiment is one important approach scientists use to gain confidence in their conclusions~\cite{mcnutt2014reproducibility}.

%Reproducibility,  rigor,  transparency,  and  independent  verification  are  cornerstones  of  the  scientific method. Of course, just because a result is reproducible does not necessarily make it  right,  and  just  because  it  is  not  reproducible  does  not  necessarily  make  it  wrong.  A  transparent  and  rigorous  approach,  however,  can  almost  always  shine  a  light  on  issues  of  reproducibility~\cite{mcnutt2014journals}.

%this might give us more idea
%https://www.nih.gov/research-training/rigor-reproducibility/principles-guidelines-reporting-preclinical-research

%Reproducibility helps to transparency.

 Over time, the scientific community has put forth various guidelines and recommendations for conducting reproducible research \cite{naturechecklist,sandve2013tenrule,wilkinson2016fair,alston2020beginnerguide,samuel2021understanding}.
Based on the current literature, we develop a set of criteria that affects the reproducibility of Knowledge Graph construction. Here, we present them as our suggested guidelines in the context of reproducibility of the construction of Knowledge Graphs, as follows:
\begin{itemize}
	\item \textbf{Availability of code and data}: One of the essential requirements for ensuring reproducible research is the availability of code and data used for constructing the KG.
	This is one of the key requirements for conducting reproducible research~\cite{sandve2013tenrule,alston2020beginnerguide}. This rule is applied to all computational research \cite{raff2019,pimental2019,samuel2023computational}. So, for the reproducible research, public access to scripts, runs, and results should be provided.
	The data used for generating KG should be available or accessible for querying.
	%In order to construct Knowledge Graph, it is important that in addition to the code, the data used for Knowledge Graph is also accessible.
	To construct a knowledge graph, not only the code but also the data should be accessible.
	Therefore, the published papers should deposit data in public repositories where available and link data bi-directionally to the published paper. Data and code shared on personal websites are considered accessible as long as the websites are maintained~\cite{alston2020beginnerguide}.

	\item \textbf{Code License}: The code used for KG construction should be accompanied by an appropriate license for reuse or reproduction. Since we found no particular mention of licenses for datasets in most of the systems, we do not report about them in this paper.

	\item \textbf{DOI for code and data}: To ensure findability, the code and data should have persistent identifiers \cite{samuel2021understanding}.
	The materials used for KG construction should be findable and linked to the published research with a permanent Digital Object Identifier (DOI). Archiving data in online repositories is one way to ensure the findability of the code and data.

	\item \textbf{Availability of execution environment}: The execution environment should be available in any format such as configuration, setup, yaml, or requirement files.
	The format for the execution environment can vary based on the programming language used for the construction of KG.
	For example, for Python, the execution environment is generally addressed by defining dependencies in standard files like requirements.txt, setup.py, and pipfile \cite{pimental2019,samuel2023computational}.
	%see page 7 section 2.1.3  Deployment Provenance from Pimentel, J. F., Freire, J., Murta, L., & Braganholo, V. (2019). A survey on collecting, managing, and analyzing provenance from scripts. ACM Computing Surveys (CSUR), 52(3), 1-38.
	According to~\cite{Pimentel2019largescale}, the lack of versions of imported libraries may cause incompatibilities and prevent the usage in other systems.
	Hence, the libraries and their version used are important information for the reproducibility of KGs.

	\item \textbf{Run instruction}: Comprehensive instructions for running the code should be provided.
	In order to reproduce the results, it is important to document the process. For computational experiments, the process of generating the results is often provided through instructions in a format like README files in the code repositories.

	\item \textbf{Online demo}:  It is desirable to have the KG itself available for use through an online demo. However, this criterion does not directly impact the reproducibility of KG systems.

	\item \textbf{SPARQL endpoint}: Having a SPARQL Endpoint to access and query the data within the Knowledge Graph offers significant advantages.

	\item \textbf{Successful regeneration}:  The code should be executable, allowing successful regeneration of the KG. %Sometimes, the code is running on the local machine of the developer well. But the public version cannot be successfully run for other researchers as the public or local version might be different.

	%\item *the used libraries are accessible. Sometimes, the special code works with a set of library which later that one does not exist anymore, or the new version of that library is not compatible with code.

	%\item *Existing of all required packages and library via e.g., -npm install or pom file

	%\item *demonstrating excellence in transparency and instill confidence in the result

	%\item The transparency of the dependent and independent variables are clear and controlled.
	\item \textbf{Provenance information}: Provenance plays a key role in the reproducibility of results. Provenance support can be used to maintain, analyze, and debug evolving knowledge graphs \cite{Belhajjame2023Online}. Both prospective and retrospective provenance offer insights into the steps required and the events that happened during the development of knowledge graphs. This information includes the addition, deletion, and updation of RDF statements \cite{Avgoustaki2016Provenance} in the construction of KGs. Additionally, it includes details regarding dataset versions, code, libraries, modules, SPARQL endpoints, etc.
\end{itemize}

Table~\ref{table:reproducibility} shows the comparison between KGs in terms of the mentioned reproducibility criteria.  Our experiments yield the following findings:

\begin{itemize}
	\item KGs such as MDKG and CKGG are not reproducible because, despite their code being publicly accessible, the necessary data for constructing these specific knowledge graphs remains inaccessible.

	\item FarseBase, MDKG, CROssBAR, and ETKG do not provide run instructions on their repository. Although their code is publicly available, it requires extra expertise to be familiar with that system to make their systems run. %Without our extra effort to try them,
    Therefore, we cannot assert their reproducibility.

	%\item CKGG failed to regenerate due to different errors due to missing data and loading data to the KG.

	\item Reproducing RTX-KG2 is challenging due to its high computational requirements. Currently, we lack resources with system specifications comparable to those of RTX-KG2. Therefore, we cannot draw any conclusions regarding its reproducibility at this time. %RTX-KG2 is hard to reproduce because of high computational requirements.

	\item Ozymandias was regenerated successfully.

\end{itemize}

%Note that, CKGG has DOI for ** but not for the code. Thus, we did not consider it.

In the RTX-KG2 repository, the authors provide links to all 70 original data sources used in its construction. However, we cannot conclude that the data of RTX-KG2 has a DOI, as some of those data sources do not have a DOI.
Moreover, a read-only endpoint\footnote{\url{http://kg2endpoint.rtx.ai:7474}} for RTX KG2 as a graph database was not available at the time of our access. % (03.06.2022).
Further demo pages were not found. Thus, we marked it with ``-'' in column 7 of Table~\ref{table:reproducibility}.

FarseBase derives its source data from Wikipedia articles composed in the Farsi language. While the repository linked with it contains the code for acquiring the source data, the actual data is not included. Since downloading the source data may not yield identical results to the data utilized in generating FarseBase, we cannot conclude whether that data is available. %mark the availability of the source data for this KG.

% Please add the following required packages to your document preamble:
% \usepackage[normalem]{ulem}
% \useunder{\uline}{\ul}{}
\begin{table*}[!htp]
	\centering
	\begin{minipage}{\textwidth}
		\caption{Comparing KGs in terms of reproducbility criteria.}
		\begin{adjustbox}{width=1\textwidth}
			\small
			\begin{tabular}{|l|l|l|l|l|l|l|l|l|l|l|}
				\hline
				\rowcolor{gray!15} & \multicolumn{3}{c|}{\textbf{Code}}                                                 & \multicolumn{2}{c|}{\textbf{Data}}    & \textbf{Online} & \textbf{SPARQL}  & \textbf{Execution} & \textbf{Run} & \textbf{Successful} \\ \cline{2-6}
				\rowcolor{gray!15} \multirow{-2}{*}{\textbf{Name}}                     & \textbf{Availability}                                              & \textbf{License} & \textbf{doi} & \textbf{Availability}  & \textbf{doi} &   \textbf{demo}    &    \textbf{endpoint}                              & \textbf{environment}                                                 &  \textbf{instruction}       & \textbf{regenerating}                                          \\ [0.5ex]	\Xhline{3\arrayrulewidth}

				CKGG & Yes\footnote{\url{https://github.com/nju-websoft/CKGG}} & No & No & No & No & Yes\footnote{\url{http://ws.nju.edu.cn/CKGG/1.0/demo}} & No & No & Yes & No \\ \hline

				\rowcolor{gray!15}CROssBAR-KG          & Yes\footnote{\url{https://github.com/cansyl/CROssBAR}}         & Yes     & No  & Yes               & Yes & Yes\footnote{\url{https://crossbar.kansil.org/}}         &  Yes                           & No                                               & No                               & No                                         \\ \hline

				ETKGCN                & Yes\footnote{\url{https://github.com/xcwujie123/Hainan\_KG}}    & No      & No  & Yes          & No      & No  & No                                                                 & No                                               & No                               & No                                          \\ \hline

				\rowcolor{gray!15}FarsBase  & Yes\footnote{\url{https://github.com/IUST-DMLab/wiki-extractor}} & No & No & - & - & Yes\footnote{\url{http://farsbase.net/sparql}} & Yes & Yes & No & No \\ \hline

				GAKG& Yes\footnote{\url{https://github.com/davendw49/gakg}} & Yes & No & No &No & Yes\footnote{\url{https://gakg.acemap.info/}} & Yes\footnote{\url{https://www.acekg.cn/sparql}} &No & Yes &No \\ \hline

				\rowcolor{gray!15}MDKG                  & Yes\footnote{\url{https://github.com/ccszbd/MDKG}}             & No      & No  & No              & No  & No                                                        & No                               & No                                               & No                               & No                                          \\ \hline

				Ozymandias             & Yes\footnote{\url{https://github.com/rdmpage/ozymandias-demo}} %\cite{ozymandiasurl}
				& Yes     & No  & Yes             & Yes & Yes\footnote{\url{https://ozymandias-demo.herokuapp.com/}} & Yes                              & No                                               & Yes                              & Yes                                         \\ \hline

				\rowcolor{gray!15}RTX-KG2 & Yes\footnote{\url{https://github.com/RTXteam/RTX-KG2}} & Yes & No & Yes & - & - & Yes\footnote{\url{https://arax.ncats.io/api/rtxkg2/v1.2/openapi.json}} & Yes & Yes & - \\ \hline

			\end{tabular}
			\label{table:reproducibility}
		\end{adjustbox}
	\end{minipage}
\end{table*}

%\subsection{Jupyter Notebook or another title}
%We conducted replications of X developed KGs where their material are available. @Sheeba: Let's here write about that we re-run the existing KG in the notebook and provide a link for them.

\section{Discussion} \label{sec:discussion}
%TODO: can we add something about iknow? or cite to our CS4Biodiversity paper
From our comparison in terms of general criteria (Table~\ref{table:comparison}), we can conclude that:
\begin{itemize}
	\item Within our dataset, the fields of medicine, biomedicine, and healthcare, which are encompassed within the broader realm of medical science, stand out as the most prevalent domains for Knowledge Graphs (KGs). This prominence can likely be attributed to the substantial volume of available data within this domain and the numerous applications that make use of KGs. Out of the 250 selected papers focusing on KGs, 56 of them (comprising 39 from medical, 12 from biomedical, and 5 from healthcare domains) account for approximately 22\% of the total (refer to Table~\ref{table:keywordSearch}).
    There is a growing trend in constructing KGs for geographic and education domains.
	%There is a trend in utilizing KGs in the biodiversity domain. %Extensive researches in the biodiversity domain have resulted in substantial data, especially tabular databases.  %TODO: can we add here also about iknow or iDiv?

	\item Most existing KGs are built based on textual data (publication) and different data sources. Interestingly, there were no KGs in our selected ones that target the tabular data. However, there is a trend to build KGs based on the tabular data.
	The Semantic Web Challenge on Tabular Data to Knowledge Graph Matching~\cite{jimenez2020semtab} is held annually to understand the semantic structure and meaning of tabular data.
	%TODO: can we add here about iknow, since in iknow we are building KG for the tabular data.

	\item Although the heuristic approaches are used to build some KGs, the machine learning approaches are the most popular construction method. %One of the difficult challenges in building a KG is to detect the relations between the entities. For this reason, machine learning approaches can help for this detection. Otherwise, human interaction needs to be involved. %I am not sure whether this argument is correct because also, in ML approaches, the training dataset will create by people manually.

	\item Although reasoning capabilities can help discover additional relationships, most KGs do not explicitly mention their use of reasoning.
	% Due to the difficulty of the reasoning, most KGs did not use or include the reasoning.

	\item KGs are widely regarded as one of the most promising ways to link information in the age of Big Data. According to the linked open data (LOD) principles~\cite{bizer2009emerging}, each knowledge resource on the web receives a stable, unique and resolvable identifier. Because of the unique identifiers, KGs can be interlinked. % with cross-domain KGs like Wikidata or other domain-specific KGs.
	However, most KGs did not provide cross-linkage. Three KGs out of eight provide the cross-link (see Table~\ref{table:comparison}). %It contributes to the creation of the Linked Open Data (LOD) dataset by interlinking to cross-domain and specific-domain KGs.

	\item The evaluation of KGs remains a challenge in this domain, as it requires the establishment of benchmarks, which is a laborious and time-consuming task.
    %Evaluation of KGs is one of the existing challenges in this domain. %TODO: citation is required? or another sentence?
	%This mainly requires a benchmark, which is an effortful and laborious task.
    Although the criteria introduced in~\cite{farber2018linked} can partially be applied in this context, KGs' evaluation seeks its own specific strategy. %I want to say, KGs evaluation needs much more study

	\item Constructing domain-specific Knowledge Graphs (KGs) using open-source code has gained popularity in recent years. As illustrated in Table~\ref{table:comparison}, all the studied platforms were developed recently. %between the years 2019 and 2021. %Building domain-specific KGs with open-source codes has become popular in recent years: eight papers related to 2019 till 2021. \sheeba{Sentence needs more structure}
\end{itemize}

%what is common? what is the difference? What is the most challenging part of each KG? What is the gap?

Following this general comparison of the studied KGs, this section explores a detailed discussion about the reproducibility test we have conducted.
To carry out this test, we examined the repository of each studied KG (as listed in Table~\ref{table:comparison}) and carefully followed the provided instructions, if available, to run the system. Note that more than one person has tested each system to ensure the reliability of the results. We draw our findings as:
%\sheeba{There is no mention of the problems we faced during the reproducibility process} 

\begin{itemize}
	\item Only 3.2\% (8 out of 250) of selected KGs have publicly available source code, indicating the need for greater encouragement towards open science and sharing data and code.

	\item Only one system out of seven open-source KGs (not considering RTX-KG2) could successfully run. 
	%Note that, we do not consider RTX-KG2.
	% pass our reproducibility test.
	This shows that only 0.4\% of selected 250 KGs (14.3\% of open-source KGs) are reproducible. This finding opens a new door for further research. %This also indicates that only public source codes do not represent the reproducibility of KGs. Available run instruction or execution environment has also a great impact on reproducibility.
	It also indicates that the availability of open-source code alone does not guarantee the reproducibility of KGs. The availability of run instructions and the execution environment also have a significant impact on reproducibility.

	%\item Another aspect is that every code before publishing should be rechecked and re-tested. Sometimes the code works well on a system (because of some specific local setting), but once the code is published online, different errors can occur without testing on another system. This issue caused 3 out of 12 KGs to fail the reproducibility test.

	\item Tracking provenance of KG construction is rarely addressed in most papers, indicating a potential gap in this aspect.

	\item Only publishing the code cannot conclude the system's reproducibility.
     It is essential to provide the code along with detailed run instructions and information about the required execution environment to facilitate reproducibility.
    %Publishing code along with the run instruction and execution environment could be forward toward reproducibility.

	\item Access to the data on which a KG is built presents another challenge for reproducibility. But, mostly domain-specific KGs are built within a project or an organization, where their data is not publicly available.

	\item It is worth mentioning that the usage of the code and data will require the corresponding licenses and considering their usage restriction.

\end{itemize}

%TODO: The main question that this paper wants to answer is whether the workflow of one of the existing domain-specific KG can be used for building a new KG in another or the same domain.
%TODO: what is the answer to this question?

%TODO: can we add something about provenance tracking?

\section{Conclusion \& Future work} \label{sec:conclusion}
Domain-specific knowledge graphs (KGs) have gained popularity due to their usability in different applications. However, the process of KG development is often challenging and time-consuming. Thus, their reproducibility can facilitate the usage of KGs in various applications. In this paper, we have conducted an analysis of existing domain-specific KGs across 19 domains, focusing on their reproducibility aspects. Our study reveals that only 0.4\% (1 out of 250) of the published domain-specific KGs is reproducible.

%Although the main focus of this study was the reproducibility of the existing domain-specific KGs, the findable, accessible, and interoperable aspects of the FAIR principle~\cite{wilkinson2016fair} can be an interesting direction for future research.\sheeba{This sentence needs to be removed from conclusion and added somewhere in introduction.}

An important future direction involves assessing the extent to which KGs effectively record their provenance. The process of maintaining KGs in alignment with their data sources can be made effortless through the establishment of a comprehensive record of source code, input data, methods, and results. This not only allows other scientists to reproduce the results, but also enables the seamless re-execution of workflows with modified input data, ensuring that KGs remain synchronized with evolving data sources. 
%And finally, by making transparent how a KG was created (open-access code, data, result, and process), we increase trust in the information provided and support open data and open science.

%TODO: @Sheeba: can we have provenance tracking aspect in this paper, or is it better we let it our for future work?

\printcredits

\section*{Declaration of competing interest}
The authors declare that they have no known competing financial interests or personal relationships that could have appeared to influence the work reported in this paper.

\section*{Acknowledgements}
SB's work has been funded by the iKNOW Flexpool project of iDiv, the German Centre for Integrative Biodiversity Research, funded by DFG (Project number 202548816).
SS's work has been funded by the Carl Zeiss Foundation for the financial support of the project ``A Virtual Werkstatt for Digitization in the Sciences (K3)'' within the scope of the program line ``Break-throughs: Exploring Intelligent Systems for Digitization - explore the basics, use applications''.
We also thank Badr El Haouni, Erik Kleinsteuber, and Anirudh Kumbakunam Ashok for testing the systems.
%+ Thanks to MScJ + Werkstatt + IMPULSE.

%\section*{Abbreviations }
%\noindent KG = Knowledge Graph \\
%can we delete this section?

\theendnotes

%% Loading bibliography style file
%\bibliographystyle{model1-num-names}
%\bibliographystyle{cas-model2-names} %main one
\bibliographystyle{ieeetr} %this style order citations by appearance
%\bibliographystyle{ieeetr}
%%\bibliographystyle{unsrtnat}

% Loading bibliography database
\bibliography{main}

%\vskip5pt

%\bio{}
%Author biography without author photo.
%\endbio

\begin{comment}

\bio{Figures/samira.jpg}
Samira Babalou, Postdoctoral researcher at Friedrich Schiller University Jena, Germany. Interested in Semantic Web, Knowledge Graphs, and Ontology Matching.
\endbio

\bigskip
\bigskip
\bigskip

\bio{Figures/sheeba.jpg}
Sheeba Samuel, Postdoctoral researcher at Friedrich Schiller University Jena, Germany. Interested in Reproducible Research, Data provenance, Scientific Data Management, Knowledge Graphs, Machine Learning.
\endbio

\bigskip
\bigskip
\bigskip

\bio{Figures/BKR.jpg}
Birgitta K\"{o}nig-Ries holds the Heinz Nixdorf  Chair for Distributed Information Systems at Friedrich Schiller University Jena, Germany.
\endbio

\end{comment}

\end{sloppypar}

\end{document}